\documentclass[preprint,showpacs,preprintnumbers,amsmath,amssymb]{revtex4}

\usepackage{graphicx}
\usepackage{dcolumn}
\usepackage{bm}

\begin{document}

\title{Gadolinium Iron Garnet as a Solid State Material for an Electron Electric Dipole Moment Search}

\author{S. K. Lamoreaux and S. Eckel}

\affiliation{Yale University, Department of Physics, P.O. Box 208120, New Haven, CT 06520-8120}

\date{Feb. 12,  2007}

\begin{abstract}
The possibility of a solid state electron electric dipole moment (EDM) experiment based on Gadolinium Iron Garnet (GdIG) is investigated.  GdIG appears to exhibit superparamagnetism and this effect can be used to enhance the electric-field-induced EDM signal at relatively a high temperatures, as compared to a simple paramagnetic system such as Gadolinium Gallium Garnet.  The sensitivity of a GdIG based experiment might be large enough that an improvement by over three orders of magnitude, compared to the existing electron EDM limit, could be obtained at a modest temperature of 4 K,  assuming that the superparamagnetic effect is not suppressed at this temperature, and that the spin relaxation time remains sufficiently rapid.
\end{abstract}
 \maketitle

\section{Introduction}

Following the original suggestion of Shapiro \cite{shap}, we have been investigating the possibility of an improved electron EDM measurement by use of paramagnetic insulating solids together with modern magnetometry \cite{lam}.  The basic idea is that when a paramagnetic insulating solid sample is subjected to an electric field, if the constituent atoms or ions of the solid have an electric dipole moment, the atom or ions spins will tend to become spin-polarized.  Because the atoms or ions carry a magnetic moment, the sample acquires a net magnetization.  We previously identified Gadolinium Gallium Garnet (GdGG) as possibly the best paramagnetic system for this study, and have done preliminary measurements indicating that the anticipated sensitivity can be achieved, assuming that spin-glass effects do not enter at temperatures of order 10 mK \cite{cyliu}. Unfortunately, such effects appear to become important near  0.1 K\  \cite{schiff},  and this will limit the noise and ultimate sensitivity of a GdGG based experiment.  We are continuing to investigate the ultimate sensitivity of a GdGG based experiment, however, according to the data in \cite{schiff}, the magnetic susceptibility has a peak near 0.1 K.

We also have performed an experiment based on the converse effect: When a ferromagnetic sample is magnetized, an electric polarization develops within it when the constituent ions or atoms carry an EDM\cite{hend} . To understand this experiment fully, the initial (zero field) magnetic permeability of the materials used in that work (Gadolinium Yttrium Iron Garnets) needs to be known.  In a literature search for the initial permeability of these materials, a single reference was found \cite{mcduff}, and an interesting effect is evident in the data presented there for GdIG.  Shown in Fig. 1 are the values for the real part of the relative permeability $\mu'$ (from $\mu=\mu'+i\mu''$)  multiplied by the measurement temperature and plotted against temperature. The linearity of the plot is striking, and shows a paramagnetic behavior.  This is rare for ferrites and ferromagnetic materials; usually, the susceptibility falls rapidly with temperature and this fact led us to consider pure paramagnetic systems such as GdGG.  An earlier solid state experiment that employed a nickel zinc ferrite suffered from this loss in susceptibility at low temperature.\cite{vasil}

For solid state electron EDM experiment that employs an applied electric field, materials with as 
as large of a magnetic susceptibility as possible are desired. This is because the EDM-induced  magnetic field signal is (mks units)
\begin{equation}
B = \chi d_e E^*/\mu_a
\end{equation}
where $\chi$ is the magnetic susceptibility, $d_e$ is the electron EDM, $E^*$ is the effective electric field (net electric
field at the lattice site of interest times the relativistic atomic or ionic enhancement factor), and $\mu_a$ is the magnetic moment 
of the paramagnetic atoms or ions that carry the electron EDM. We see immediately
that $\chi$ should be as large as possible. Note that the sample temperature does not explicitly
enter into this formula. We are therefore free to use whatever means possible to increase
$\chi$. As can be seen from the data in \cite{mcduff}, the relative permeability of GdIG is $\mu'= \chi + 1 = 58$ at 77 K; assuming the paramagnetic $1/T$ behavior holds for GdGG, this
is already equal to the permeability of GdGG at 40 mK. If the trend in relative permeability shown in Fig. 1 continues to sufficiently temperature (4 K), a sensitive experiment, using GdIG instead of GdGG, will be possible.  Scaling the results of \cite{lam} to a relative initial permeability of 1000 (four times that of GdGG at 10 mK), with sample area 10 cm$^2$, indicates an experimental sensitivity of $2\times 10^{-27}\ e{\rm cm}\sqrt{\rm sec}$, with 10 kV/cm applied to the sample at a modest temperature of 6 K. Thus, in one second of measurement, a sensitivity at the level of the best experimental limit for the electron EDM \cite{demille} can be achieved.

\section{Previous Work}

There have been results from three solid state electron EDM experiments reported.  The first of these, mentioned above, used a nickel-zinc ferrite, with $\mu'=11$ at 4 K. Our recent work employed GdGG, and obtained a better value than \cite{vasil}, but not as large as we had anticipated.  This is due in part to the wrong values of the electric field being used in \cite{vasil,cyliu}.  

The internal electric field at a site on a cubic lattice site is given by \cite{kittel} as $\epsilon E_0/3$ where $\epsilon$ is the dielectric constant, and $E_0$ is the externally applied field strength, i.e., the field between the sample field plates without the dielectric material.  However, the analysis presented in \cite{kittel} is applicable only when the frequency of the applied field is much higher than the ion vibration frequency or principal phonon modes as there is an implicit assumption that the ion position does not change with application of the field.  Clearly, these are optical frequencies, and the results must be modified for a static field. This problem was carefully treated in \cite{sush1} where the change in equilibrium position due to the applied field is taken into account, see Sec. VI of that paper. A more refined calculation is presented in \cite{sush2}, and when modified by a more refined relativistic enhancement factor \cite{sush3}, the energy shift of a Gd$^{+3}$ ion, due to an electron EDM of $1.6\times 10^{-26}\ e\ $cm (the present experimental limit \cite{demille}) is
\begin{equation}
\Delta \varepsilon = 3.8\times 10^{-22}\ {\rm eV}
\end{equation}
for an internal electric field of $E=10$ kV/cm (externally applied field $E_0=\epsilon E$). It should be noted that for this calculation, $\epsilon\approx 30$ was used, whereas we found $\epsilon=19$ for both single crystal and polycrystalline samples, and that the electron EDM energy shift is proportional to a factor (see \cite{sush1}, Eq. (44)),
\begin{equation}
X={(\epsilon-1) E \over 32 \pi e n_{\rm Gd}}={(\epsilon -1 ) E_0\over \epsilon} {1\over 32 \pi e n_{\rm Gd}},
\end{equation}
and in the limit of large $\epsilon$, the sensitivity is independent of $\epsilon$.  We can calculate the apparent electric field at a Gd$^{+3}$ ion site, by taking the ionic enhancement factor as 2.2 from \cite{sush3}, 
\begin{equation}
{3.8\times 10^{-22}\ {\rm eV}\over 30}=2.2\times\ 1.6\times 10^{-27}\ e{\rm cm} {E_{apparent}\over E_0}\ \rightarrow\ E_{apparent}=3.6\ {\rm kV/cm}
\end{equation}
when $E_0=10$ kV/cm, so the field is reduced by a factor of three at the ion site compared to the applied field.  Interestingly, this is the field expected within a spherical cavity in a dielectric material, for large values of dielectric constant.  Clearly, the value $E=\epsilon E_0/3$ used in \cite{vasil} is probably not correct, although the correction in that case is relatively small because $\epsilon=2.2$ for the nickel-zinc ferrite used in that work.

Furthermore, when using a different value for the dielectric constant for different GdGG samples, the factor $X$ discussed above must be modified. As mentioned already, this factor makes the experimental sensitivity independent of $\epsilon$ for $\epsilon>>1$.  This means that the limit obtained in relation to the efforts \cite{cyliu} are a factor of 1.5 too small, taking $\epsilon\approx 20$ compared to $\epsilon=30$ in \cite{sush1}, without modifying $X$.  Also, a density correction has to be applied also that varies between samples.

\section{Superparamagnetism: Gadolinium Iron Garnet}

Superparamagnetism exists in a ferromagnetic system when the size of the domains be-
comes so small that the interaction energy between domains, or the energy required for domain wall movement,
is below the thermal fluctuation energy, $k_bT$.
A magnetized ferromagnet is actually in a metastable state, and the relaxation time to the
ground state for a domain-domain interactions can be described by an Ahrrenius-type equation,
\begin{equation}
\tau = \tau_0 \exp(K_uV/k_bT)
\end{equation}
where $K_u$ is the uniaxial anistropy constant, and $V$ is the volume of a particle or domain. \cite{11}
In this context, it is useful to think of the spin of a domain as undergoing a Brownian motion
around the easy axis, with $K_u$ being the energy to change alignment from one easy axis to
another. Alternatively, we can apply the same equation to the movement of a domain wall (e.g., increase in
domain volume), with $K_uV$ replaced by an appropriate ``activation energy" associated with an increase in
domain volume, in which case
\begin{equation}
\tau=\tau_0\exp(E_d/k_bT)
\end{equation}
where $E_d$ is the energy barrier for domain wall movement. It is generally accepted that at low frequencies and low fields, the relaxation time is determined by domain wall movement.\cite{boz}
 
If the time  (e.g., the inverse of the frequency of an oscillating magnetic
field) of an experimental measurement of a magnetic property is longer than $\tau$, a ferromagnetic material can behave as a superparamagnetic system, with the properties that the susceptibility increases as $1/T$ and that there is very little magnetic hysteresis. On the other hand, for faster times, hysteresis becomes important, and the system
appears as a usual ferromagnet. Normally, $V$ is large, and $\tau$ becomes infinite for all practical
purposes. Nevertheless, there are well-known instances where V is small enough for $\tau$ to be small,
the most famous of which is the superparamagnetic limit on the density of a hard drive of
a few hundred gigabits per square inch. A smaller bit size, as determined by the thickness of the magnetic layer (to restrict the transverse field leakage distance) is unstable against thermal
fluctuations. Increasing this density is of course an area of active research.  Although not named as such, superparamagnetism was first investigated by Oliver Heaviside in  1887 \cite{heavi}.  He showed that finely powdered iron dust, ``Worked up with wax into solid cores (1 wax to 5 or 6 iron by bulk),''  showed no hysteresis effects, or, in his words, the effect of the iron in such a state is, ``Exactly equivalent to self-inductance.''

For the EDM problem, superparamagnetism is useful, as we have discussed already. As
shown in Fig. 1, we see that the real part of the relative initial permeability of GdIG can be described by
\begin{equation}
\mu'(T)= {5830\ K\over T}-17.8
\end{equation}
where the constant part probably represents the finite accuracy of the data, but does suggest
that $\mu'= 1$ at 310 K, which is well below the Curie temperature of GdIG, 590 K, but is close to the ``compensation temperature" where the three sublattice magnetization cancel \cite{kittel}.  A question remains as to the range of validity of this relation; it is probably not overly hopeful to assume that it is unmodified to temperatures approaching 4 K. It is interesting to note that in \cite{mcduff}, for GdYIG materials, the relative permeability falls with temperature. It is of course not possible to tell if this is due to the domain size, due to a different value of $K_u$, or due to domain movement. In \cite{mcduff}, the ``dispersion frequency" (peak in $\mu''$ as a function of frequency) increases for GdYIG mixtures at the temperature is lowered; this is the opposite of what is expected for superparamagnetism.  There are no dispersion frequencies listed for GdIG except for 77 K where the first dispersion frequency is given as 27 MHz.  It is impossible to know how the scaling varies with temperature based on this single value; in addition, it is likely that only a small subset of the sample spins contribute to the observed effect.  Different domains can have different $E_b$, so as the temperature is lowered, there will likely be a gradual decrease in $\tau$.

As the correlation time decreases, the noise due to spin fluctuations appear in a narrower
frequency range, and therefore have a larger contribution near zero frequency, where an EDM
experiment would be performed[13]. In a fundamental sense, if the relative permeability continues to increase as
$1/T$, at sufficiently low temperature, the correlation time will increase as $\exp(E_d/k_bT)$ as
will the zero frequency spin noise, and will eventually outpace the increase in $\mu'=1+\chi'$, the real part of the susceptibility.

Using the results in \cite{budker}, we can determine the effects of the relaxation time in a quantitative fashion.  The density of Gd$^{+3}$ in GdIG is $\rho_{Gd}=1.24\times 10^{22}$/cc, and the magnetic moment comes mostly from the Gd which has $\mu_a=7\mu_b$ where $\mu_b$ is the Bohr magneton.  A signal to noise of unity indicates the level with which an EDM can be discriminated above background noise.  The results in \cite{budker} can be recast to give a fundamental EDM sensitivity, in terms of the sample volume $V$,
\begin{equation}
d_e={ \mu_0 \mu_a2 \sqrt{2\rho_{Gd} T_1}   \over  \chi e E^* \sqrt{V} }
\end{equation}
so to achieve $d_e=1.6\times 10^{-27}\ e$cm in one second of averaging, for $E^*=10 {\rm kV/cm}$, $V=10$ cc, and $\chi=1000$ requires that $T_1<10^{-10}$ sec, or a relaxation frequency of about 1 GHz.  On the other hand, we can ask what level of sensitivity can be achieved at 77 K, with $\chi=57$ and $T_1=1/(2\pi\times 27 {\rm MHz})=5\times 10^{-9}$ sec and $E^*=10$ kV/cm:  $d_e=2\times 10^{-25}\ e{\rm cm}\sqrt{\rm sec}$, so after ten days of averaging, a limit of $2\times 10^{-28}\ e$cm can be achieved.  Therefore a substantial improvement in the experimental limit appears as imminently feasible if even cooling below 50 K or so is not possible.

Of course, GdIG can be permanently magnetized when a large enough magnetic field is applied.
The arguments above assume that the applied fields are very small. For GdIG, the super-
paramagnetism is a non-linear effect, in that it disappears above some critical field, of value
to be determined. An earlier study of the susceptibility of GdIG has a large applied field so
the superparamagnetic effect is not evident.\cite{wolf}  More recent studies \cite{gdig} of the magnetic properties of GdIG show that the susceptibility remains relatively large at 5 K.  However, the slope, hence inferred susceptibility, at zero field in Fig. 9 of \cite{gdig} is determined by the sample geometry; for a spherical sample in a solenoid, the effect of the ``demagnetizing field"  limits the apparent susceptibility to $\chi=M/B_0=3$ in the limit of large $\chi$ (or $\mu$) (see Eq. (5.115) of \cite{jackson}), which is close to inferred value of $\chi$ for all of the curves in Fig. 9. 

The maximum field before a sample might be magnetized significantly and permanently can be estimated by analogy with usual ferromagnetic materials, which have paramagnetic properties at low fields while in the demagnetized state.  The sample begins to acquire a significant magnetization in an external field when the applied field is larger than (neglecting the geometry-dependent demagnetizing field)
\begin{equation}
B_0={\mu_0 M_s\over \chi}
\end{equation}
where $M_s$ is the  saturation magnetization.  If we take $\chi=1000$, $M_s=\mu_a \rho_{Gd}$, we find that $B_0=10$ Gauss.  

\section{Possible Experiment}

As shown in Fig. 2, operation of a solid state EDM experiment at temperature near 4 K
will allow the use of a ``warm"-bore system as heat loads are not so much of an issue. The heat
shields and dewar will have to be made of materials with low electrical conductivity (plastics,
G-10), and their thermal conductivity could be improved by incorporating thin wires which
would not contribute as much Johnson noise as continuous metal sheets. The entire system
could be easily contained in a superconducting shield.

The magnetic field due to an EDM of $d_e=1.6\times 10^{-27}\ e$cm, with $E^*=10$ kV/cm,  $\chi=1000$ is
\begin{equation}
B={\chi d_eE^*\over \mu_a}\approx 0.05 {\rm \ fT}.
\end{equation}
By placing a superconducting pickup loop with inductance $L_p$ around the sample, the flux $\Phi=BA=L_p I$ can be converted to a current.  This current can be measured in two ways, either by placing another loop around magnetometer cell, or by sending the current to a SQUID magnetometer.  

For the first option, the present best magnetometer noise has been achieve in a spin-exchange-relaxation-free (SERF) potassium  (K) magnetometer, at the level of 0.1 fT$/\sqrt{\rm Hz}$ \cite{all}.  If the sample is relatively large compared to the magnetometer, it will be possible to have a smaller loop around the magnetometer and step up the field at the magnetometer.  The optimal transfer of signal occurs when the magnetometer coil inductance matches the pickup coil inductance.  An order of magnitude increase in the field magnitude at the magnetometer is not beyond comprehension.    

A simpler approach is to use a SQUID magnetometer.  If we treat the SQUID as a current measuring device, its sensitivity is determined by the fractional flux quantum noise, typically $1\mu\Phi_0/\sqrt{\rm Hz}$ for commercial grade SQUIDs, and an order of magnitude better for research units.  The input current is converted to a SQUID flux through a coupling coil's mutual inductance, which is typically $0.2\ \mu{\rm A}/\Phi_0$ \cite{qd1}, where $\Phi_0$ is the superconductor flux quantum.  Taking a sample area of 10 cm$^2$, using $B=0.1$ fT, and $L_P=5\ \mu$H (which is about a factor of two larger than the SQUID input inductance), a current $I_p=20$ fA is generated, implying a SQUID signal of  $0.1\ \mu\Phi_0/\sqrt{\rm Hz}$, a factor of 10 below the intrinsic SQUID sensitivity . It is unlikely that use of a transformer to better match $L_p$ to the input inductance would help as most any pickup loop will be within an order of magnitude of the SQUID input inductance.\cite{qd2}

We see immediately that the magnetometer method offers at least a factor of 10 better signal to noise that the SQUID method.  The magnetometer method can be further improved by use of a flux concentrator loop, as mentioned above. The possibility of such an improvement when using a conventional magnetometer was already mentioned in \cite{lam}. On the other hand, the simplicity of the SQUID system is quite attractive, and the sensitivity of the magnetometer configuration is possibly better than the fluctuation noise of the sample.  If this latter point proves correct, the added complexity of an atomic magnetometer is unwarranted.  

An even simpler experimental concept is shown in Fig. 3.  Here, two disks of GdIG sandwich a magnetometer, and although the loss in signal is at least a factor of five for 1 cm thick disks, separated by 3 cm, the sensitivity is comparable to the SQUID method outlined above.

\section{Required Studies}

There are a number of issues that need to be addressed to evaluate the efficacy of GdIG
for a solid state electron EDM experiment. The following is a listing, not necessarily in
order of importance or complete, of points to be addressed:

\noindent 
1. Determine the effective electric field $E^*$ for GdIG which includes both the atomic
enhancement and the dielectric shielding. There is no reason to suspect that there
should be surprises compared to previous calculations of GdGG.

\noindent
2. Calculate the effective electric field susceptibility, noting that the electric field
couples significantly to only the Gd$^{+3}$ spins.  Most of the magnetization of GdIG comes from Gd$^{+3}$ at temperatures below 77 K.

\noindent
3. Verify the $1/T$ scaling of the permeability to low temperature, and measure $\mu''$  
as a function of frequency and temperature. The latter should give the most sensitive measure of the
domain relaxation time, and will provide a means to determine the minimum operating temperature. These are the most important parameters in regard to the design of
an experiment.  It could be that only a small fraction of the domains contribute to $\mu''$ so the scaling in \cite{budker} must be modified; the value of $T_1$ is no longer simply determined.  Also the issue of domain rotation vs. growth needs to be studied. If domain rotation is the principal contribution to the relaxation time, then the results in \cite{budker} again need to be modified.

\noindent
4. Establish the maximum magnetic field that can be applied before the sample begins to
acquire a permanent magnetization. This is related to the low-frequency hysteresis,
which is important regarding permanent magnetization due to displacement, currents,
microdischarges, and leakage currents.

\noindent
5. Determine the effects of sample preparation on the superparamagnetic properties.
As the crystalline structure of the sample becomes coarser, it might be expected that
the superparamagnetic effect will be suppressed, due to the $V$ factor in the exponent
of the Arrhenius time, or an increase in the energy required for domain wall movement.

\noindent
6. Determine the degree to which residual sample magnetization can be eliminated. The limiting residual
magnetism will place limitations on the types of
magnetometry that can be used. Also the stability to the residual magnetization is
important. Given that the Curie temperature of GdIG is 590K, it
should be possible to anneal a sample while it is placed in a magnetic shield and heated within an oven.  In addition, the compensastion temmperature of GdIG is near room temperature, so careful 
handling and avoidance of exposure to large magnetic fields would be necessary to ensure that the
residual magnetization remains low. A crucial related point is addressed in (4) above.

\section{Conclusion}

GdIG exhibits the behavior normally associated with a superparamagnetic system, so
it is probably fair to refer to GdIG as having this property. Unfortunately, the initial
permeabilities of the rare earth garnets have not been extensively studied, and a significant
amount of effort will be required to assess the suitability of GdIG for a solid state electron
EDM experiment.  Nevertheless, GdIG has potential to provide the basis
of an experiment that can achieve a truly significant improvement in the electron EDM limit.  If it is possible to achieve $\mu'\approx 1000$ for GdIG by cooling to 4 K, after ten days of averaging, an EDM limit at the level of $10^{-29}\ e{\rm cm}$ can be achieved using a commercial SQUID magnetometer.  Furthermore,
GdIG will be useful as a cryogenic ferrite and will certainly find applications in cryogenic
electronic systems. In particular, if the permeability remains large at 4 K, GdIG will be very
useful in the construction of low noise amplifiers, including DC parametric amplifiers.

As a final note, there are possible applications to other fundamental measurements.  By searching for a GdIG magnetization correlated with velocity or direction in absolute space, Lorentz violating interactions can be sought, with perhaps to six orders of magnitude improvement over existing limits.

\section{Acknowledgements}

I thank Prof. D. Budker for critical comments on the manuscript, and A.O. Sushkov for pointing out the importance of the domain structure in determining the spin noise.

\section{Appendix: Detuning of the Gradiometer Condition due to Sample Permeability}

In our work\cite{cyliu}, we employed a superconducting pickup loop, operated in a concentric loop planar gradiometer configuration. Unfortunately, the permeability of the sample alters the pickup loop inductances, and spoils the gradiometer tuning, and reduces  the experiment sensitivity slightly because the loop inductance is increased. In the case of a single loop or radius $a$ around an infinitely long cylinder of permeability $\mu$ and radius $b$, the increase in inductance is \cite{smythe},
\begin{equation}
\delta L= 2\mu_0 a2\int_0^\infty \Phi(k)[K_1(ka)]2 dk
\end{equation}
where $\mu_0$ is the permeability of free space, $K_1(ka)$ is the first order modified Bessel function, and
\begin{equation}
\Phi(k)= {(\mu -1)kb I_0(kb)I_1(kb)\over (\mu -1)kb K_0(kb)I_1(kb)+1}\approx {I_0(kb)\over K_0(kb)}
\end{equation}
where $I_0$ and $K_0$ are the zero-order modified Bessel functions, and the approximation holds when $\mu>>1$. The effect of finite length is small, provided that length $\ell \geq 2b$, which is the case for the experiment.
In the case $a=b$ for a loop directly on a cylinder, $\delta L\approx 6\mu_0 a\approx 0.1\ \mu$H, for $a=1.2$ cm, a 20\% effect.  When $a=\sqrt{2} b$, $\delta L=0.02\ \mu$H, a much smaller effect.  The loss in sensitivity is around 10\%, leaving the principal problem due to this effect as the detuning of the gradiometer condition, which is temperature dependent, and of order 10\%.  Usually, gradiometers are fabricated to provide 1\% common mode rejection.  We see a substantial limitation to the degree of common mode rejection possible when there is a permeable sample in the central loop.

\begin{figure}
\begin{center}
\includegraphics[
width=4in ] {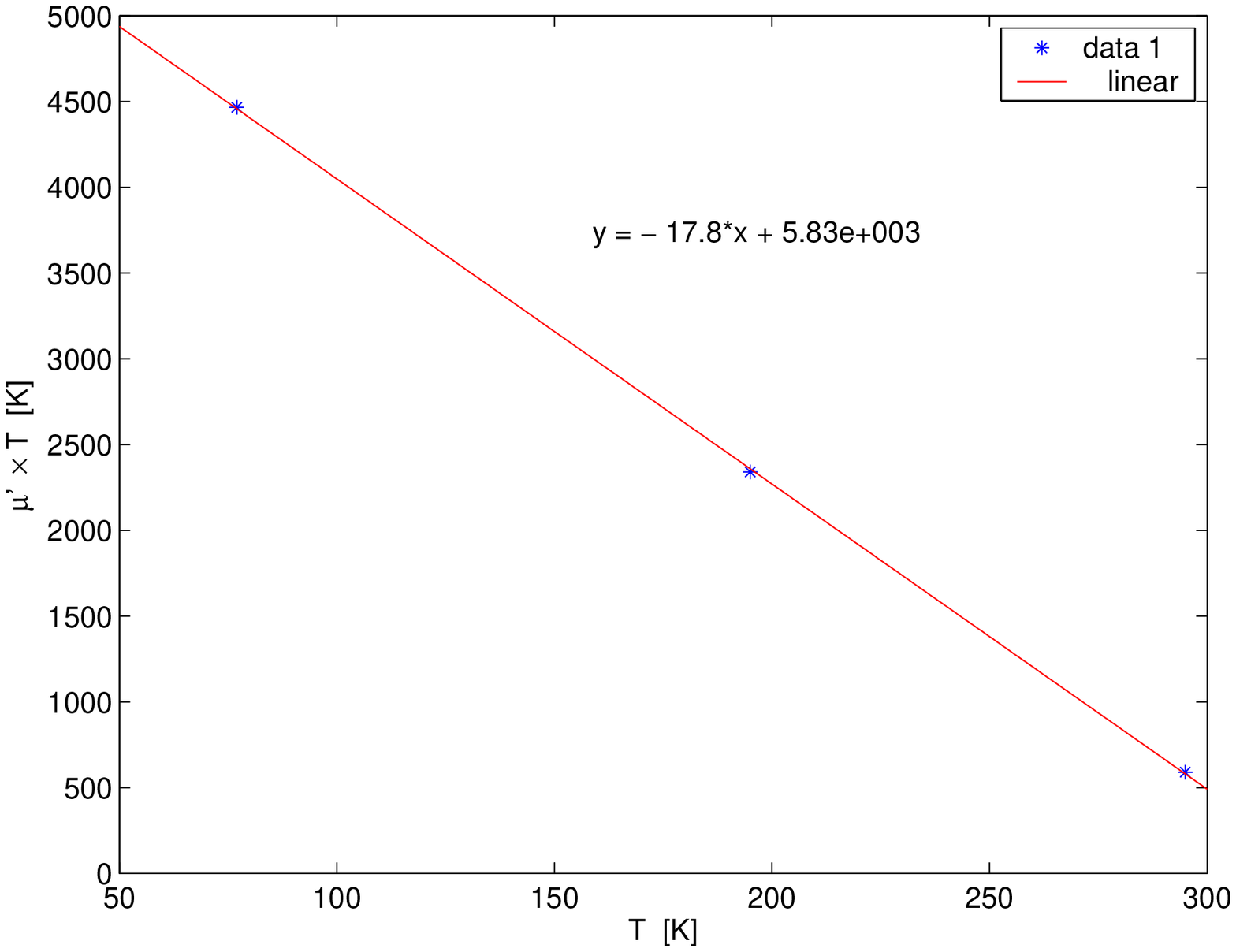} \caption{The values of GdIG relative initial permeability taken from \cite{mcduff}, and plotted as $T\ \mu'(T)$ vs. $T$. The linear form shows that $\mu'\propto 1/T+$ constant.}
\end{center}
\end{figure}

\begin{figure}
\begin{center}
\includegraphics[
width=4in ] {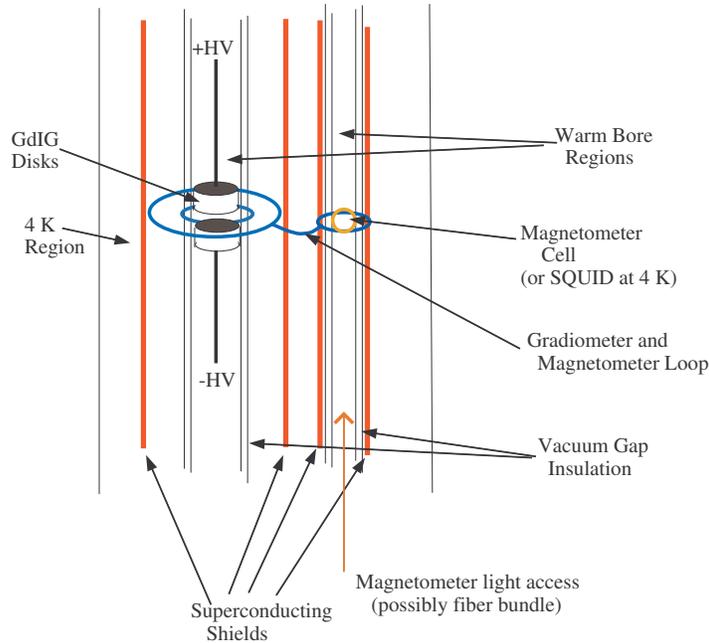} \caption{Schematic of a cryostat with a warm bore that incorporates a SERF magnetometer. Two disks of GdIG have opposite sign high voltage applied to electrodes on the disk faces, with a ground plane iplaced between the disks.  A SQUID magnetometer could be incorporated instead, in which case a second warm bore is not required.  If it turns out that the magnetic properties of GdIG remain favorable to 4 K (or below), the warm bores can be dispensed with if a SQUID is employed. However, as pointed out in the text, better sensitivity can be achieved with an atomic magnetometer, in which case a single warm bore would be required, with the GdIG held at 4 K. }
\end{center}
\end{figure}

\begin{figure}
\begin{center}
\includegraphics[
width=4in ] {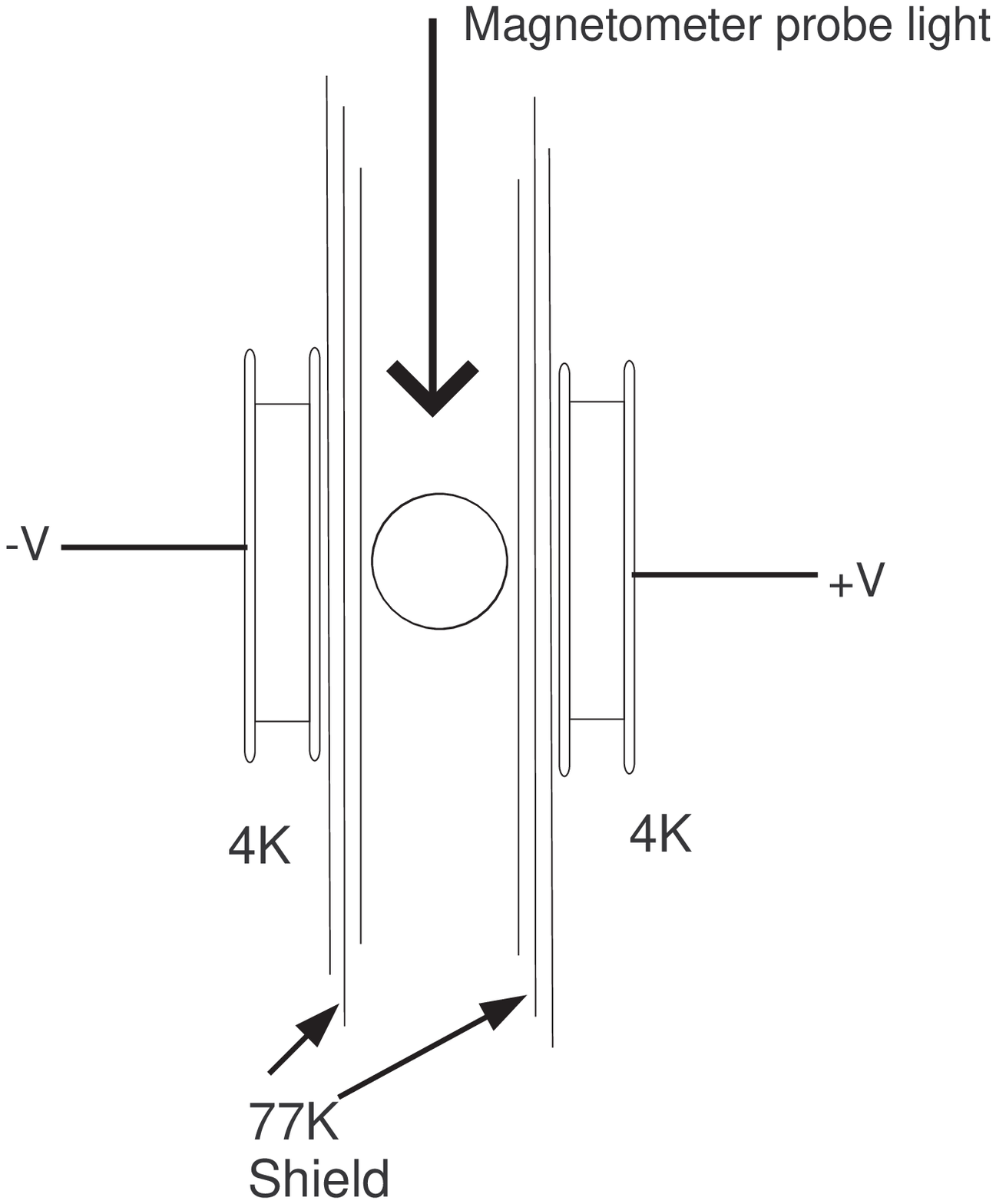} \caption{Schematic of a cryostat with a warm bore that incorporates a SERF magnetometer between two disks of GdIG that have opposite sign high voltage applied  and a ground plane is placed between the disks.  For disk separation larger than the thickness of the disks, there is a substantial loss in magnetic field signal at the magnetometer location, compared to the internal field of the GdIG samples.}
\end{center}
\end{figure}


\begin{thebibliography}{99}



\bibitem{shap} F.L. Shapiro, Sov. Phys. Uspekhi 95, 145 (1968).
\bibitem{lam} S.K. Lamoreaux, Phys. Rev. A (2002).
\bibitem{cyliu} C.-Y. Liu and S.K. Lamoreaux, Mod. Phys. Lett A 19, 1235 (2004).
\bibitem{schiff} P. Schiffer et al., Phys. Rev. Lett. 74, 2379 (1995).
\bibitem{hend} B.J. Heindenreich et al., Phys. Rev. Lett. 95, 253004 (2005).
\bibitem{mcduff} G.E. McDuffie, Jr., J.R. Cunningham, Jr., and E.E. Anderson, Jour. App. Phys., Supplement
to Vol. 31, 47S (1960).
\bibitem{vasil} B.V. Vasil'ev and E.V. Kolycheva, Sov. Phys. JETP 47, 243 (1978).
\bibitem{demille} B.C. Regan, E.D. Commins, C.J. Schmidt, and D. DeMille,
Phys. Rev. Lett. {\bf 88}, 071805 (2002).
\bibitem{kittel} Charles Kittel, {\it Introduction to Solid State Physics, 4th Ed.} (Wiley, New York, 1971).
\bibitem{boz} Richard M. Bozorth, {\it Ferromagnetism} (Van Nostrand, Princeton, N.J., 1951).  See, e.g., Chap. 18., where it is pointed out that the usual mechanism for magnetization is by domain movement.
\bibitem{sush1} S. Kuenzi, O.P. Sushkov, V.A. Dzuba, and J.M. Cadogan,
Phys. Rev. A 66, 032111 (2002).
\bibitem{sush2} T.N. Muckhamedjanov, V.A. Dzuba, and O.P. Sushkov, Phys. Rev. A {\bf 68}, 042103 (2003).
\bibitem{sush3}  V.A. Dzuba, O.P. Sushkov, W.R. Johnson, and U.I. Safranova, Phys. Rev. A {\bf 66}, 032105 (2002).
\bibitem{11} L. N\'eel, Ann. Geophys. C.N.R.S. 5, 99 (1949).
\bibitem{heavi} O. Heaviside, {\it The Electrician}, Sept. 25, 1885, p. 375. (Reprinted in {\it Electrical Papers by Oliver Heaviside, vol. II}, p. 159 (Macmillan and Co., London, 1892).)
\bibitem{wolf} W.P. Wolf and R.M. Bozorth, Phys. Rev. 124, 449 (1961).
\bibitem{gdig} T. Yamagishi et al., Philosophical Magazine,
{\bf 85}, 1819 (2005).
\bibitem{jackson} J.D. Jackson, {\it Classical Electrodynamics, 2nd Ed.} (Wiley, New York, 1975).
\bibitem{budker} D. Budker et al., Phys. Rev. A 73, 022107 (2006).
\bibitem{all} J.C. Allred et al, Phys. Rev. Lett. 89, 130801 (2002).
\bibitem{qd1} Quantum Design, Technical Specification for DC SQUID.
\bibitem{qd2} Quantum Design, SQUID Application Note 1052-202.
\bibitem{smythe} W.R. Smythe, {\it Static and Dynamic Electricity} (McGraw-Hill, New York, 1950);  Chap. 8, prob. 15.  
\end{thebibliography}
\end{document}